\documentstyle[aps]{revtex}
\input{psfig}  
\begin{document}
\vspace{2.in}
\title{Nonvanishing zero modes in the light-front current} 
\author{ Ho-Meoyng Choi and Chueng-Ryong Ji }
\address{
Department of Physics,
North Carolina State University,
Raleigh, N.C. 27695-8202}
\maketitle
\begin{abstract}
We find that the zero mode($q^{+}=0$ mode of a continuum theory)
contribution is crucial to obtain the 
correct values of the light-front current $J^{-}$ in the
Drell-Yan($q^{+}=0$) frame. In the exactly solvable model of 
(1+1)-dimensional scalar field theory interacting with gauge fields, 
we quantify the zero mode contribution 
and observe that the zero mode effects are very large for the
light meson form factors even though they are substantially reduced
for the heavy meson cases. 
\end{abstract}
\newpage
\baselineskip=20pt
One of the distinguishing features in light-front quantization is the
rational energy-momentum dispersion relation which gives a sign 
correlation between the light-front energy($P^{-}$) and the light-front  
longitudinal momentum($P^{+}$). In the old-fashioned time-ordered perturbation
theory\cite{Brodsky}, this sign correlation allows one to remove the so-called 
``Z-graphs" such as the diagram of particle-antiparticle pair 
creation(annihilation) from(to) the vacuum. 
As an example, in the theory of scalar fields interacting with gauge 
fields\cite{GS,Saw}, the covariant triangle diagram shown in Fig.1(a) 
corresponds to only two light-front time-ordered diagrams shown in Figs.1(b)
and 1(c), while in the ordinary time-ordered perturbation theory, Fig.1(a)
would correspond to the six time-ordered diagrams including the 
``Z-graphs". Furthermore, the Drell-Yan(or $q^{+}=0$) frame
may even allow one to remove the diagram shown in Fig.1(c) because of the same 
reasoning from the energy-momentum dispersion relation and the conservation
of the light-front longitudinal momenta at the vertex of the gauge field
and the two scalar fields.  
\begin{figure}
\psfig{figure=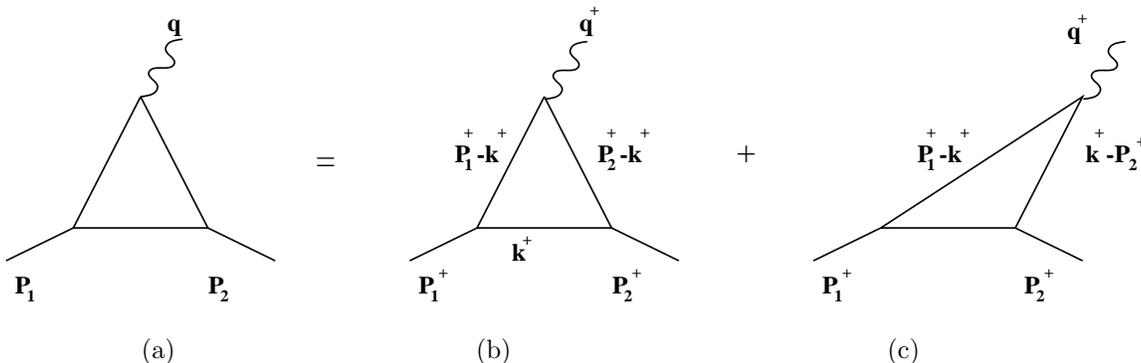,width=6in}
\hspace{1.5cm}(a)\hspace{4cm}(b)\hspace{5cm}(c)
\caption{Covariant triangle diagram (a) is represented as the sum of
light-front triangle diagram (b) and the light-front pair-creation
diagram (c).}
\end{figure}
Based on this idea, the Drell-Yan($q^{+}=0$) frame is frequently used
for the bound-state form factor calculations.
Taking advantage of $q^{+}=0$ frame, one may need to 
consider only the valence diagram shown in Fig.1(b), where the three-point
scalar vertices should be replaced by the light-front bound-state 
wavefunction. Successful description of various hadron form factors 
can be found in the recent literatures\cite{Dziem,CRJ,HC,Jaus,Card} 
using the light-front quark model.

In this paper, however, we point out that even at $q^{+}=0$ frame
one should not overlook the possibility of non-zero contribution from the 
non-valence( pair creation or annihilation) diagram shown in Fig.1(c). 
As we will show explicitly in the simple $(1+1)$-dimensional scalar field 
theory interacting with gauge fields, the current $J^{-}$ is not immune to
the zero mode contribution shown in Fig.1(c) at $q^{+}=0$. While  
the current $J^{+}$ does not have any zero mode contribution from 
Fig.1(c), the processes that involve more than one form factor, $e.g.$, 
semileptonic decay processes, require the calculations of more components
of the current other than $J^{+}$ in order to find all the necessary 
form factors in $q^{+}=0$ frame. 
For instance, in the analysis of the semileptonic decays between 
two pseudoscalar mesons, two form factors, $f_{\pm}(q^{2})$, are involved
and one has to use not only $J^{+}$ but also $J^{-}$(or $J^{\perp}$  
in $3+1$ dimensions) to obtain both form factors in $q^{+}=0$ frame. 
Thus, the zero mode contribution is crucial to obtaining the correct results
of electroweak form factors.
Only a brief exactly solvable model calculation is provided here. A full,
detailed treatment of $(3+1)$ dimensional semileptonic decay processes 
such as $K\to\pi$, $B\to\pi$, $B\to D$ etc. will be presented in a 
separate communication.   
We first describe the general formalism of the semileptonic decay form 
factors for non-zero momentum transfer in ($1+1$)-dimensions and then 
discuss the zero mode problem in the limiting cases of the form factors 
as $q^{+}\to 0$. 

The semileptonic decay of a $Q_{1}\bar{q}$ bound state into
another $Q_{2}\bar{q}$ bound state is governed by the weak current, viz., 
\begin{eqnarray}
J^{\mu}(0)&=&<P_{2}|\bar{Q_{2}}\gamma^{\mu}Q_{1}|P_{1}>=
f_{+}(q^{2})(P_{1}+P_{2})^{\mu} + f_{-}(q^{2})(P_{1}-P_{2})^{\mu},
\end{eqnarray}
where $P_{2}=P_{1}-q$ and the non-zero momentum transfer square 
$q^{2}=q^{+}q^{-}$ is time-like, $i.e.$, $q^{2}=[0,(M_{1}-M_{2})^{2}]$.
One can easily obtain $q^{2}$ in terms of the fraction $\alpha$ as follows
\begin{eqnarray}
q^{2}&=& (1-\alpha)(M^{2}_{1}-\frac{M^{2}_{2}}{\alpha}), 
\end{eqnarray}
where $\alpha=P^{+}_{2}/P^{+}_{1}=1-q^{+}/P^{+}_{1}$. 
Accordingly, the two solutions for $\alpha$ are given by 
\begin{eqnarray}
\alpha_{\pm}&=&\frac{M_{2}}{M_{1}}\biggl[ 
\frac{ M^{2}_{1}+M^{2}_{2}-q^{2}}{2M_{1}M_{2}}
\pm \sqrt{\biggl(\frac{ M^{2}_{1}+M^{2}_{2}-q^{2}}
{2M_{1}M_{2}}\biggr)^{2}-1} \biggr]. 
\end{eqnarray}
The $+(-)$ sign in Eq.(3) corresponds to the daughter meson
recoiling in the positive(negative) $z$-direction relative to
the parent meson. At zero recoil($q^{2}=q^{2}_{\rm max}$) and
maximum recoil($q^{2}=0$), $\alpha_{\pm}$ are given by
\begin{eqnarray}
\alpha_{+}(q^{2}_{\rm max})&=& 
\alpha_{-}(q^{2}_{\rm max})=\frac{M_{2}}{M_{1}},
\nonumber\\
\alpha_{+}(0) &=& 1,\hspace{0.5cm}
\alpha_{-}(0)=\biggl(\frac{M_{2}}{M_{1}}\biggr)^{2}.  
\end{eqnarray}
In order to obtain the form factors $f_{\pm}(q^{2})$ which are
independent of $\alpha_{\pm}$, we can define    
\setcounter{equation}{0}
\renewcommand{\theequation}{\mbox{5\alph{equation}}}
\begin{eqnarray}
<P_{2}|\bar{Q_{2}}\gamma^{\mu}Q_{1}|P_{1}>|_{\alpha=\alpha_{\pm}}&\equiv&
2P_{1}^{+}H^{+}(\alpha_{\pm})\hspace{.2cm}{\rm for}\hspace{.2cm}\mu=+,\\
&\equiv& 2\biggl(\frac{M^{2}_{1}}{P^{+}_{1}}\biggr)H^{-}(\alpha_{\pm})    
\hspace{.2cm}{\rm for}\hspace{.2cm}\mu=-, 
\end{eqnarray}
and obtain from Eq.(1)
\setcounter{equation}{0}
\renewcommand{\theequation}{\mbox{6\alph{equation}}} 
\begin{eqnarray}
f_{\pm}(q^{2})&=&\pm \frac{(1\mp \alpha_{-})H^{+}(\alpha_{+}) - 
(1\mp \alpha_{+})H^{+}(\alpha_{-})}{\alpha_{+}-\alpha_{-}} 
\hspace{.2cm}{\rm for}\hspace{.2cm}\mu=+,\\ 
&=&\pm \frac{(1\mp \beta_{-})H^{-}(\alpha_{+}) -
(1\mp \beta_{+})H^{-}(\alpha_{-})}{\beta_{+}-\beta_{-}}
\hspace{.2cm}{\rm for}\hspace{.2cm}\mu=-,  
\end{eqnarray}
where $\beta_{\pm}=\alpha_{-}(0)/\alpha_{\pm}$. 

Now, the current $J^{\mu}(0)$ obtained from the covariant triangle  
diagram of Fig.1(a) is given by    
\renewcommand{\theequation}{7}
\begin{eqnarray}
J^{\mu}(0)&=& \int d^{2}k\frac{1}{ (P_{1}-k)^{2}-m^{2}_{1}+i\epsilon}
(P_{1}+P_{2}-2k)^{\mu}\frac{1}{ (P_{2}-k)^{2}-m^{2}_{2}+i\epsilon}
\frac{1}{ k^{2}-m^{2}_{\bar{q}}+i\epsilon}.  
\end{eqnarray}  
From this, we obtain for the ``$\pm$"-components of the current $J^{\mu}(0)$ as
\renewcommand{\theequation}{8} 
\begin{eqnarray} 
J^{\pm}(0)&=& -2\pi i(I^{\pm}_{1}+I^{\pm}_{2}),
\end{eqnarray}   
where $I^{\pm}_{1}$ and $I^{\pm}_{2}$ corresponding to diagrams Figs.1(b) 
and 1(c), respectively, are given by   
\setcounter{equation}{0} 
\renewcommand{\theequation}{\mbox{9\alph{equation}}} 
\begin{eqnarray}
I^{+}_{1}(\alpha)&=& \int^{\alpha}_{0}dx\frac{1-2x+\alpha}{x(1-x)(\alpha-x)
\biggl(M^{2}_{1}-\frac{m^{2}_{1}}{1-x}-\frac{m^{2}_{\bar{q}}}{x}
\biggr)\biggl(\frac{M^{2}_{2}}{\alpha}-\frac{m^{2}_{2}}{\alpha-x}-
\frac{m^{2}_{\bar{q}}}{x}\biggr )}, \\
I^{+}_{2}(\alpha)&=& \int^{1}_{\alpha}dx \frac{1-2x+\alpha}{x(1-x)(\alpha-x)
\biggl(M^{2}_{1}-\frac{m^{2}_{1}}{1-x}-\frac{m^{2}_{\bar{q}}}{x}
\biggr) \biggl(\frac{M^{2}_{2}}{\alpha}+\frac{m^{2}_{1}}{1-x}-
\frac{m^{2}_{2}}{\alpha-x}-M^{2}_{1}\biggr )},
\end{eqnarray} 
and 
\setcounter{equation}{0} 
\renewcommand{\theequation}{\mbox{10\alph{equation}}}  
\begin{eqnarray}
I^{-}_{1}(\alpha)&=& \int^{\alpha}_{0}dx \frac{M^{2}_{1}+M^{2}_{2}/\alpha 
-2m^{2}_{\bar{q}}/x}{x(1-x)(\alpha-x)
\biggl(M^{2}_{1}-\frac{m^{2}_{1}}{1-x}-\frac{m^{2}_{\bar{q}}}{x}
\biggr)\biggl(\frac{M^{2}_{2}}{\alpha}-\frac{m^{2}_{2}}{\alpha-x}-
\frac{m^{2}_{\bar{q}}}{x}\biggr)},\\
I^{-}_{2}(\alpha)&=&\int^{1}_{\alpha}dx \frac{M^{2}_{2}/\alpha -M^{2}_{1} 
+2m^{2}_{1}/(1-x)}{x(1-x)(\alpha-x)\biggl(M^{2}_{1}-\frac{m^{2}_{1}}{1-x}-
\frac{m^{2}_{\bar{q}}}{x}\biggr)\biggl(\frac{M^{2}_{2}}{\alpha}  
+ \frac{m^{2}_{1}}{1-x}- \frac{m^{2}_{2}}{\alpha-x} - M^{2}_{1}
\biggr)}. 
\end{eqnarray} 
Note that at zero momentum transfer limit, $q^{2}=q^{+}q^{-}\to 0$, 
the contributions of $I^{\pm}_{2}(\alpha)$ 
come from either $lim_{q^{+}\to0}I^{\pm}_{2}(\alpha)=
I^{\pm}_{2}(\alpha_{+}(0))$ or $lim_{q^{-}\to0}I^{\pm}_{2}(\alpha)=
I^{\pm}_{2}(\alpha_{-}(0))$.  
It is crucial to note in $q^{+}=0$ frame that while 
$I^{+}_{2}(\alpha_{+}(0))$ vanishes, $I^{-}_{2}(\alpha_{+}(0))$  
does not vanish because the integrand has a singularity even though
the region of integration shrinks to zero\cite{Hwang}.
Its nonvanishing term is thus given by
\setcounter{equation}{0}
\renewcommand{\theequation}{11} 
\begin{eqnarray}
I^{-}_{2}(\alpha_{+}(0))= -\frac{2}{m^{2}_{1}-m^{2}_{2}}{\rm ln}\biggl(
\frac{m^{2}_{2}}{m^{2}_{1}}\biggr). 
\end{eqnarray} 
This nonvanishing term is ascribed to the term proportional to 
$k^{-}=P^{-}_{1}-m^{2}_{1}/(P^{+}_{1}-k^{+})$ in Eq.(10b), which prevents 
Eq.(10b) from vanishing in the limit, $\alpha\to 1$. 
This is precisely the contribution from ``zero mode" at $q^{+}=0$ frame.  
$I^{-}_{2}(\alpha_{+}(0))$ should be distingushed from the other 
nonvanishing pair-creation diagrams at 
$q^{-}=0$ frame, i.e., $I^{\pm}_{2}(\alpha_{-}(0))$.  
Some relevant but different applications of zero modes were discussed 
in the literatures\cite{zero}.

In Table I, we summarized the form factors $f_{\pm}(0)$ 
obtained from both currents, $J^{+}$ and $J^{-}$, for different
zero momentum transfer limit, i.e., $q^{+}= 0$ or $q^{-}= 0$. 
As shown in Table I, the non-valence contributions, 
$I^{\pm}_{2}(\alpha_{\pm}(0))$, are separated from the
valence contributions, $I^{\pm}_{1}(\alpha_{\pm}(0))$.
Of special interest, we observed that the form factor $f_{-}(0)$ at 
$q^{+}= 0$ is no longer free from the zero mode, $I^{-}_{2}(\alpha_{+}(0))$.

To give some quantitative idea how much these non-valence contributions 
$I^{\pm}_{2}(\alpha_{\pm}(0))$ are for a few different decay processes,
we performed model calculations for $K\to\pi$, $B\to\pi$, and $B\to D$ 
transitions in $(1+1)$ dimensions using rather widely used constituent 
quark masses, $m_{u(d)}=0.25$ GeV, $m_{c}=1.8$ GeV, 
and $m_{b}=5.2$ GeV. Numerically, we first verified that the form factors,  
$f_{+}(0)$ and $f_{-}(0)$, obtained from the $q^{+}= 0$ frame are 
in fact exactly the same with $f_{+}(0)$ and $f_{-}(0)$ obtained from the
$q^{-}=0$ frame, respectively, once the non-valence contributions( 
including zero mode) are added. 
The non-valence contributions to the form factors
of $f_{\pm}(0)$ at $q^{-}=0$ are also shown in Table II.
In Figs.2a(b)-4a(b), the effects of pair-creation(non-valence)
diagram to the exact form factors are shown for the non-zero momentum
transfer region for the above three decay processes.
Especially, the zero mode contributions $I^{-}_{2}(\alpha_{+}(0))$
to the exact solutions for the $f_{-}(0)$ at 
$q^{+}=0$, i.e., $f^{\rm Z.M.}_{-}(0)/f^{\rm full}_{-}(0)$, are 
estimated as $6.9$ for $K\to\pi$, $0.03$ for $B\to\pi$, and $0.12$ for 
$B\to D$ decays. 
The zero mode contributions on $f_{-}(0)$ at $q^{+}=0$
frame are drastically reduced from the light-to-light meson transition to
the heavy-to-light and heavy-to-heavy ones.
This qualitative feature of zero mode effects
on different initial and final states are expected to remain same
even in $(3+1)$ dimensional case, even though the actual quantitative values
must be different from $(1+1)$ dimensional case.

Furthermore, we have found the effect of zero mode to the EM form factor;
\setcounter{equation}{0}
\renewcommand{\theequation}{12} 
\begin{eqnarray}
J^{\mu}(0)&=& (2P_{1}-q)^{\mu}F_{M}(Q^{2}). 
\end{eqnarray} 
The EM form factor at $q^{+}=0$ using $J^{-}(0)$
current is obtained by 
\setcounter{equation}{0}
\renewcommand{\theequation}{13} 
\begin{eqnarray}
F_{M}(0)&=& N\biggl\{\int^{1}_{0}dx\frac{M^{2}-m^{2}_{\bar{q}}/x}
{x(1-x)^{2}\biggl(M^{2}-\frac{m^{2}_{q}}{1-x}-
\frac{m^{2}_{q}}{x}\biggr)^{2}}+ 1/m^{2}_{q}\biggr\},
\end{eqnarray}     
where $N$ is the normalization constant and
the $1/m^{2}_{q}$ in Eq.(13) is the ``zero mode" term. Numerically, using 
the previous quark masses, the effects of zero modes on the form factors of 
$F_{\pi}(0)$ and $F_{B}(0)$, i.e., 
$F^{\rm Z.M.}_{\pi}(0)/F^{\rm full}_{\pi}(0)$ and
$F^{\rm Z.M.}_{B}(0)/F^{\rm full}_{B}(0)$, are  estimated as     
$16.9$ and $0.75$, respectively.   
Again, the zero mode contribution is drastically reduced for the heavy
meson form factor. However, it gives a very large effect on the light 
meson form factors.   
The similar observation on the EM form factor was made in the Breit 
frame recently\cite{Fred}. In ($3+1$) dimensions, however, we note that
the relation between the Breit frame and the Drell-Yan frame involves 
the transverse rotation in addition to the boost and therefore the results
obtained from the Breit frame cannot be taken as the same with those obtained 
from the Drell-Yan frame or vice versa. 

In conclusion, we investigated the zero mode effects on the form factors 
of semileptonic decays as well as the electromagnetic transition in 
the exactly solvable model. Our main observation was the nonvanishing 
zero mode contribution to the $J^{-}$ current and our results are 
directly applicable to the real $(3+1)$ dimensional calculations.  
The effect of zero mode to the $f_{-}(0)$ form factor 
is especially important in the application for the physical semileptonic 
decays in the Drell-Yan($q^{+}=0$) frame.  
To the extent that the zero modes have a significant contribution to 
some physical observables as shown in this work, one may even conjecture
that the condensation of zero modes could lead to the nontrivial 
realization of chiral symmetry breaking in the light-front quantization 
approach. The work along this line is in progress.  
 
It is a pleasure to thank Bernard Bakker, Stan Brodsky, Matthias Burkardt,
Tobias Frederico, Dae Sung Hwang and Carl Shakin for several 
informative discussions. This work was supported by the U.S. DOE under
contracts DE-FG02-96ER 40947. 

\newpage 
\begin{table}
\caption{Form factors of $f_{\pm}(0)$ obtained for different zero-momentum 
transfer limit, $q^{+}=0$ and $q^{-}=0$. The notation of $\alpha_{p(m)}$
used in table are defined as $\alpha_{p}=1+\alpha_{-}(0)$ and
$\alpha_{m}=1-\alpha_{-}(0)$, respectively. }
\begin{tabular}{|c|c|c|}
Form factor & $q^{+}=0$ & $q^{-}=0$\\ \hline
$f_{+}(0)$ & $I^{+}_{1}(\alpha_{+}(0))/2$
& $\sum_{i=1}^{2}I^{-}_{i}(\alpha_{-}(0))/2M^{2}_{1}$\\ \hline
$f_{-}(0)$ & $[\sum_{i=1}^{2}I^{-}_{i}(\alpha_{+}(0))/M^{2}_{1}
-\alpha_{p}I^{+}_{1}(\alpha_{+}(0))/2]/\alpha_{m}$
& $\sum_{i=1}^{2}[I^{+}_{i}(\alpha_{-}(0))
-\alpha_{p}I^{-}_{i}(\alpha_{-}(0))/2M^{2}_{1}]/\alpha_{m}$  \\
\end{tabular}
\end{table}
\begin{table}
\caption{ Zero-mode(Z.M.) and non-valence(N.V.) contributions to the exact form 
factors of $f_{\pm}(0)$ for the semileptonic decays of $K(B)\to\pi$ and $B\to D$
in $(1+1)$ dimensions. 
We distinguished the zero mode contribution at $q^{+}=0$ from the usual
non-valence one at $q^{-}=0$.}
\begin{tabular}{|c|c|c|c|c|c|}
Frame& Ratio of $f^{\rm N.V.(Z.M.)}_{\pm}(0)$ to $f^{\rm full}_{\pm}(0)$ 
& N.V.(Z.M.) factor & $K\to\pi$ 
& $B\to\pi$ & $B\to D$\\ \hline
$q^{+}=0$& $f^{\rm Z.M.}_{+}(0)/f^{\rm full}_{+}(0)$ & None & 0 & 0 & 0 \\ 
\cline{2-6} 
& $f^{\rm Z.M.}_{-}(0)/f^{\rm full}_{-}(0)$ &$\propto I^{-}_{2}(\alpha_{+}(0))$ 
& 6.9 & 0.03 & 0.1 \\ \hline
$q^{-}=0$& $f^{\rm N.V}_{+}(0)/f^{\rm full}_{+}(0)$ & 
$\propto I^{-}_{2}(\alpha_{-}(0))$ & 2.8& 1.3& 0.05 \\ \cline{2-6}
& $f^{\rm N.V1.}_{-}(0)/f^{\rm full}_{-}(0)^{[a]}$ & 
$\propto I^{+}_{2}(\alpha_{-}(0))$  & 3.8 & 3.8 & 0.6 \\ \cline{2-6}
& $f^{\rm N.V2.}_{-}(0)/f^{\rm full}_{-}(0)^{[a]}$ & 
$\propto I^{-}_{2}(\alpha_{-}(0))$ & $-11.1$ & $-4.0$ & $-1.1$ \\  
\end{tabular}
\end{table}
$^{[a]}$ We show the separate contributions of the non-valence 
terms proportional to 
$I^{+}_{2}(\alpha_{-}(0))$ and $I^{-}_{2}(\alpha_{-}(0))$ to the 
exact form factor of $f_{-}(0)$ at $q^{-}=0$.  
\newpage
\setcounter{figure}{0}
\renewcommand{\thefigure}{\mbox{2\alph{figure}}}
\begin{figure}
\hspace{2cm}\psfig{figure=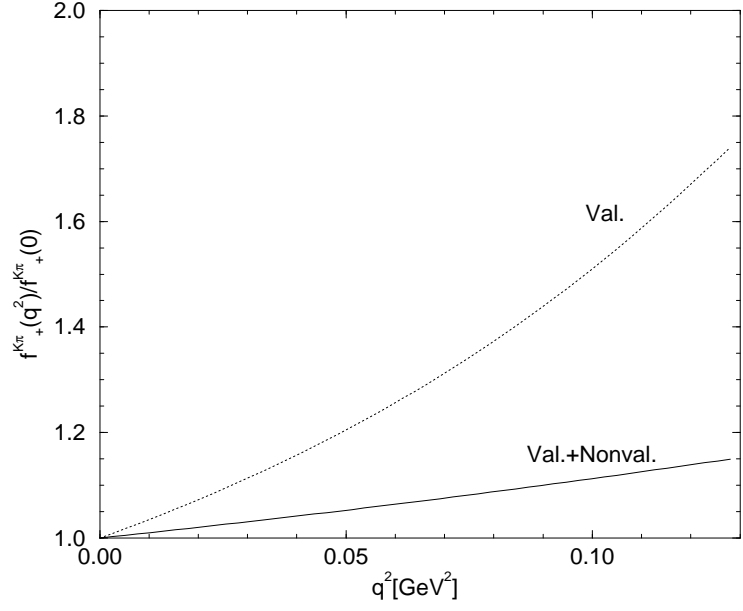,width=4.5in}
\caption{Normalized form factor of $f_{+}(q^{2})$ for $K\to\pi$
in $(1+1)$ dimension. The solid line is the result from the  
valence plus non-valence contributions. 
The dotted line is the result from the valence contribution.}
\end{figure}
\begin{figure}
\hspace{2cm}\psfig{figure=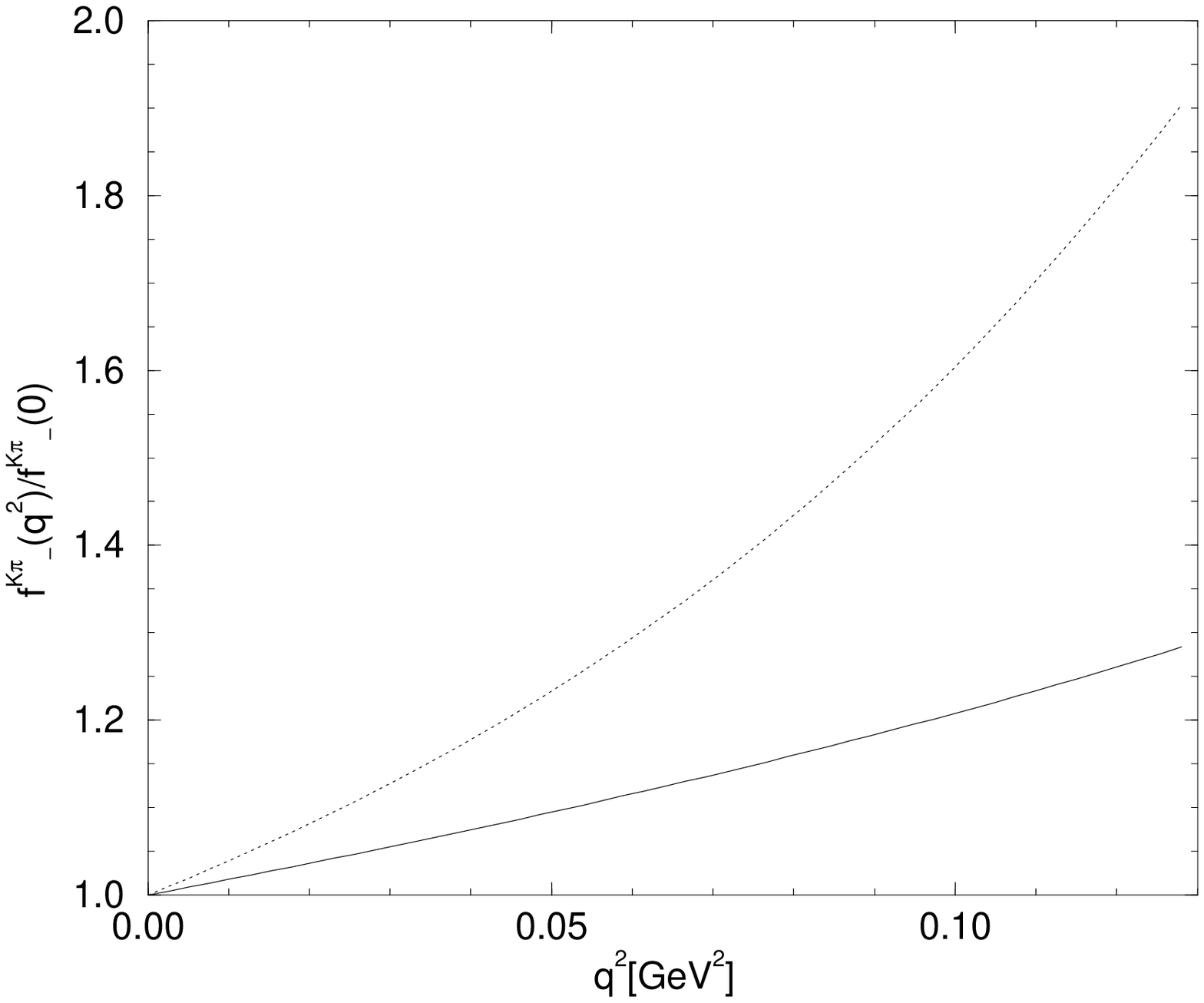,width=4.5in}
\caption{Normalized form factor of $f_{-}(q^{2})$ for $K\to\pi$
in $(1+1)$ dimension. The same line code as in Fig.2a is used.}
\end{figure}
\setcounter{figure}{0}
\renewcommand{\thefigure}{\mbox{3\alph{figure}}}
\begin{figure}
\hspace{2cm}\psfig{figure=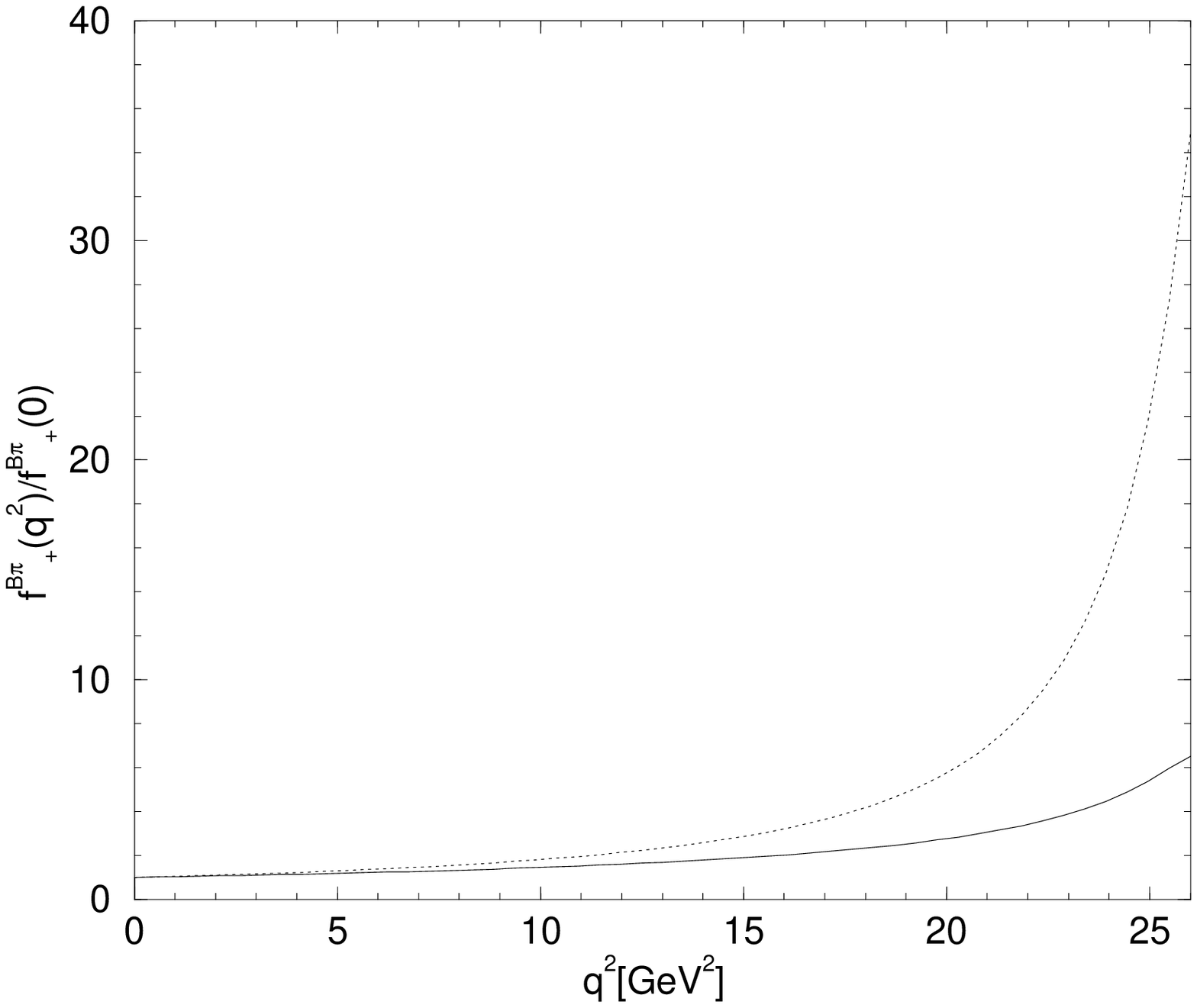,width=4.5in}
\caption{Normalized form factor of $f_{+}(q^{2})$ for $B\to\pi$
in $(1+1)$ dimension. The same line code as in Fig.2a is used.}
\end{figure}
\begin{figure}
\hspace{2cm}\psfig{figure=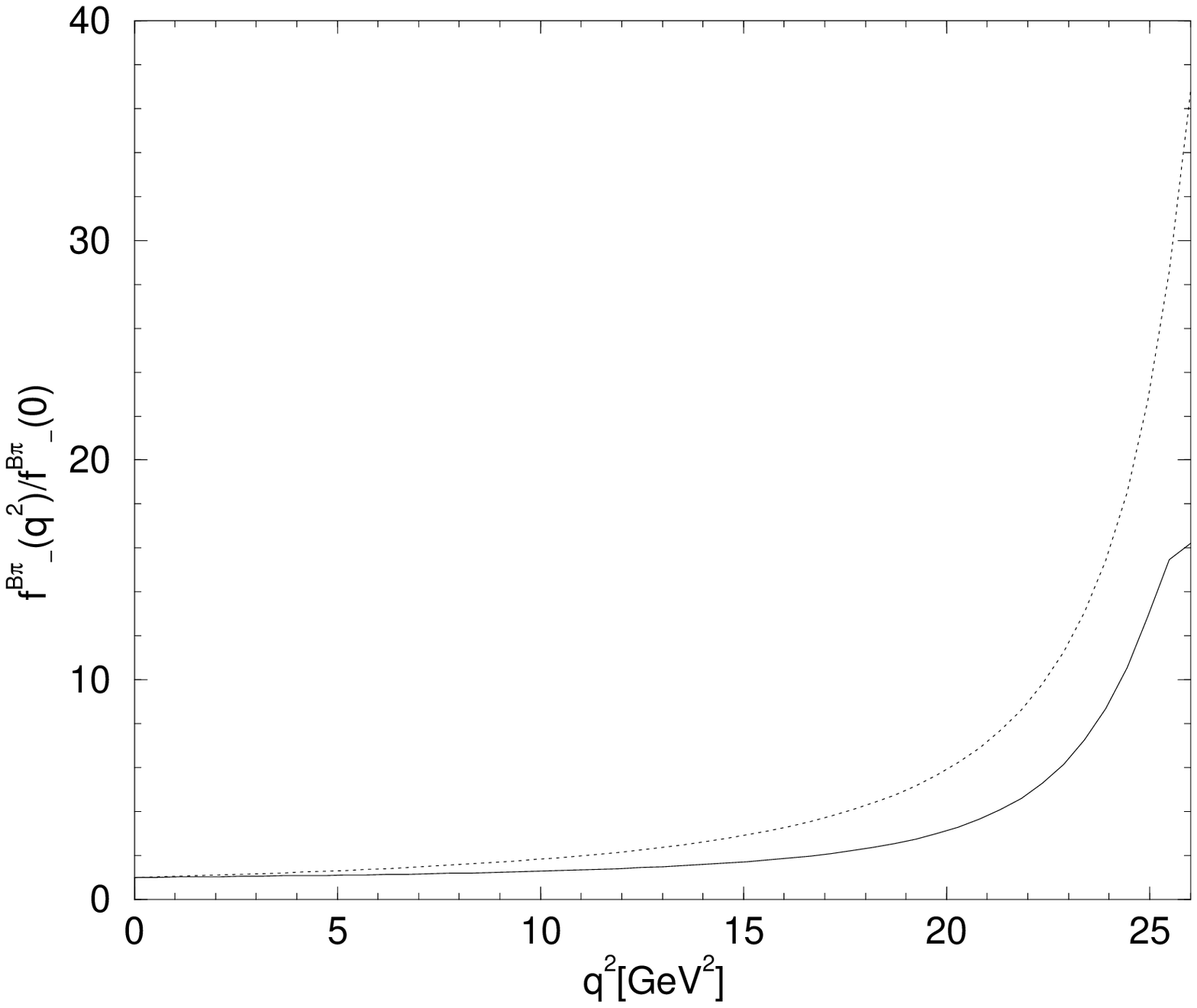,width=4.5in}
\caption{Normalized form factor of $f_{-}(q^{2})$ for $B\to\pi$
in $(1+1)$ dimension. The same line code as in Fig.2a is used.}
\end{figure}
\setcounter{figure}{0}
\renewcommand{\thefigure}{\mbox{4\alph{figure}}}
\begin{figure}
\hspace{2cm}\psfig{figure=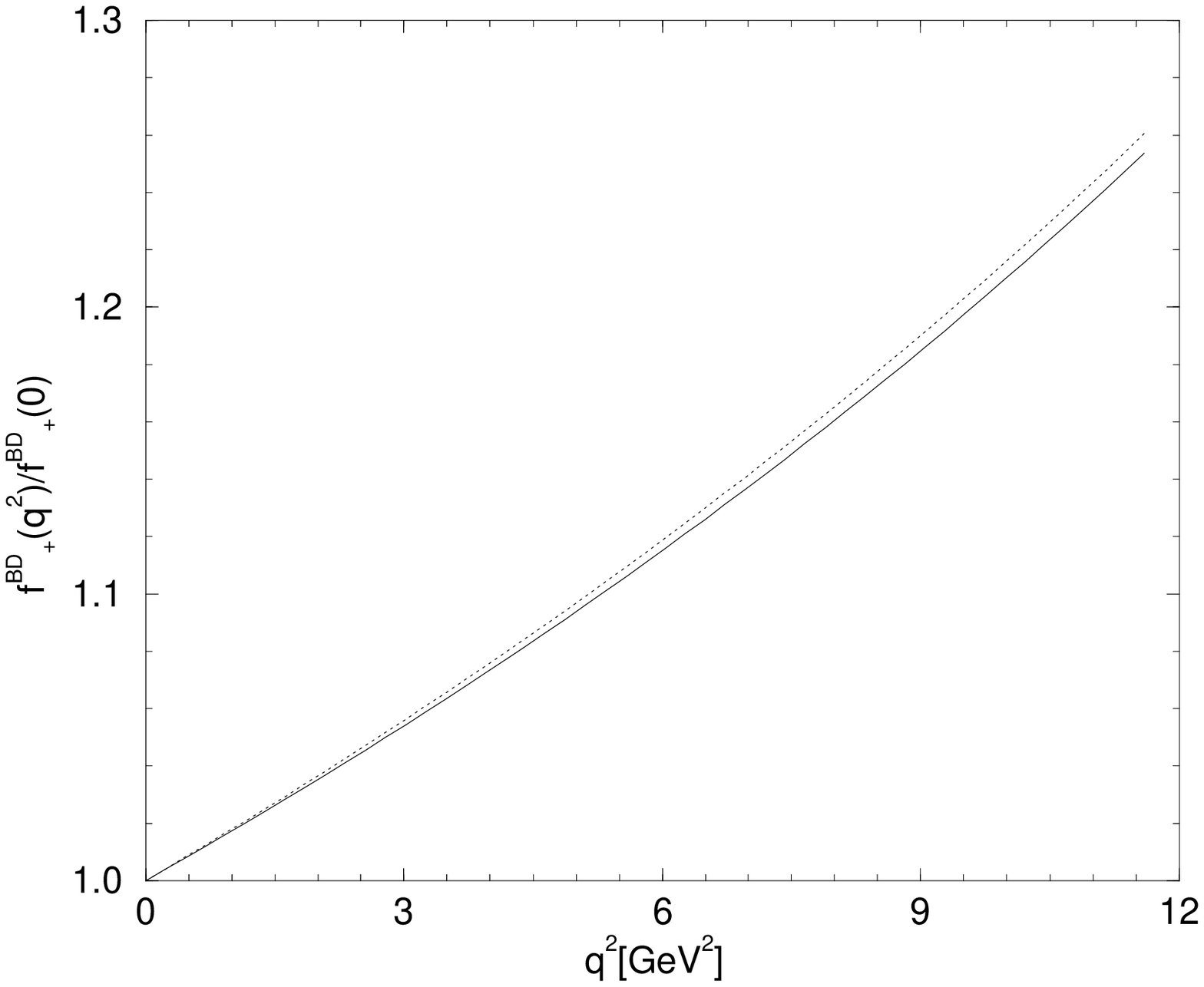,width=4.5in}
\caption{Normalized form factor of $f_{+}(q^{2})$ for $B\to D$
in $(1+1)$ dimension. The same line code as in Fig.2a is used.}
\end{figure}
\begin{figure}
\hspace{2cm}\psfig{figure=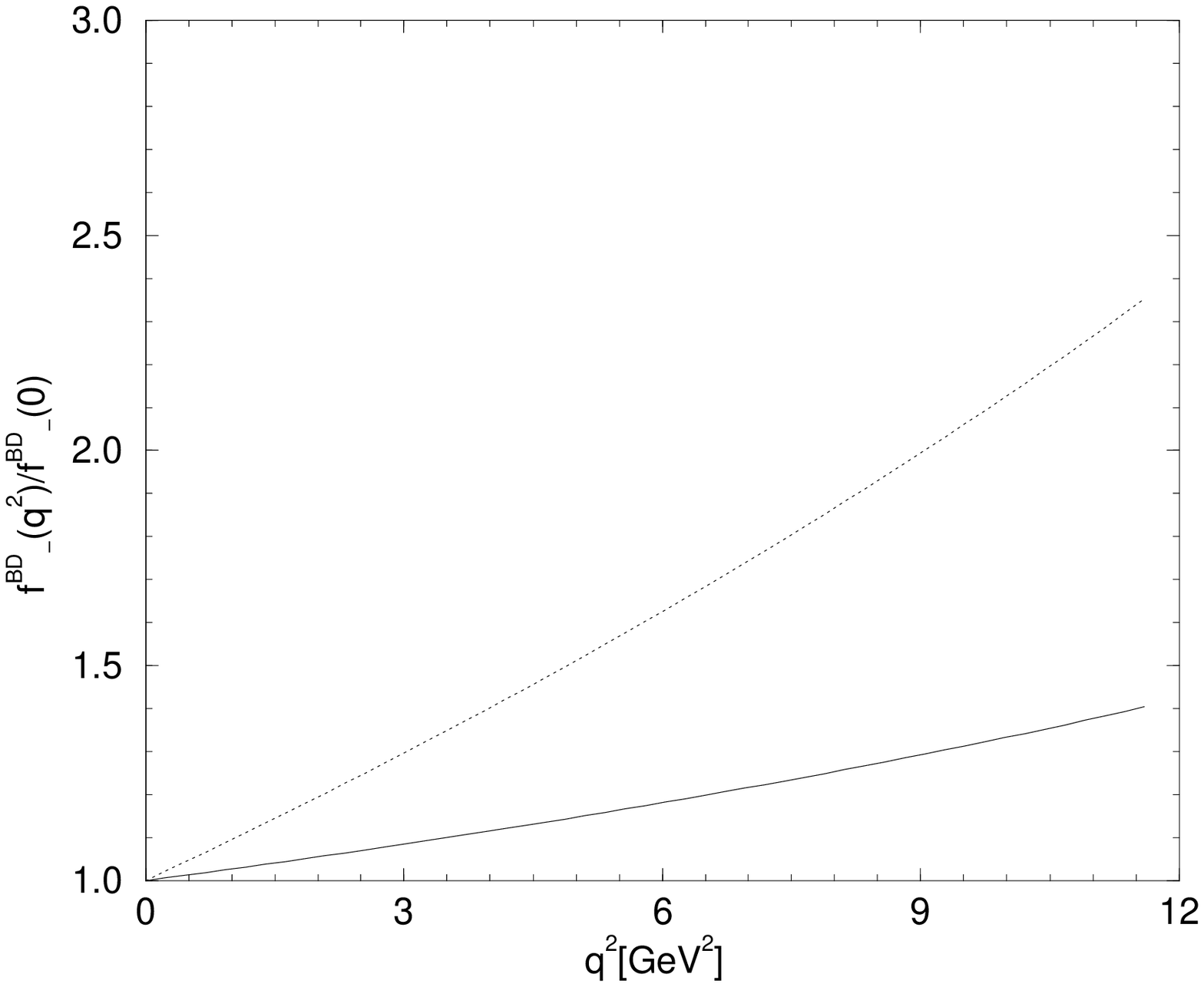,width=4.5in}
\caption{Normalized form factor of $f_{-}(q^{2})$ for $B\to D$
in $(1+1)$ dimension. The same line code as in Fig.2a is used.}
\end{figure}
\end{document}